\documentstyle[preprint,aps]{revtex}
\newcommand{\beq}{\begin{equation}}
\newcommand{\eeq}{\end{equation}}
\newcommand{\eq}[1]{eq.(\ref{#1})}
\begin{document}
\draft
\preprint{PSU/TH/181; hep-ph/9703400}
\tighten
\title {Gyromagnetic Ratios of Bound Particles}
\author {Michael I. Eides \thanks{E-mail address:  
eides@phys.psu.edu, eides@lnpi.spb.su}}
\address{ Department of Physics, Pennsylvania 
State University, 
University Park, PA 16802, USA\thanks{Temporary address.}\\ 
and
Petersburg Nuclear Physics Institute,
Gatchina, St.Petersburg 188350, Russia\thanks{Permanent address.}}
\author{Howard Grotch\thanks{E-mail address: h1g@psuvm.psu.edu}}
\address{Department of Physics, Pennsylvania State University,
University Park, PA 16802, USA}
\date{March, 1997}

\maketitle
\begin{abstract}
A new approach to calculation of the binding corrections 
to the magnetic moments of the constituents in a loosely bound system, 
based on the Bargmann-Michel-Telegdi equation, is suggested. Binding 
corrections are calculated in this framework, and the results confirm 
earlier calculations performed by other methods.  Our method clearly 
demonstrates independence of the binding corrections on the magnitude of the 
spin of the constituents.  
\end{abstract} 
%\pacs{PACS numbers: }

\section{Introduction}

It is a common practice to describe the Zeeman effect in atoms in terms of 
the gyromagnetic ratios of the constituents, and also to use when needed the 
Clebsch-Gordon coefficients. However, due to the binding effects, magnetic 
moments of the electrons and the nuclei do not coincide exactly with the 
magnetic moments of the free particles. Respective bound-state corrections 
to the gyromagnetic ratios of the constituents in the hydrogenic atoms have 
been calculated a long time ago in Refs.\cite{gh,fa,co}. Explicit 
calculations in these works were done for the case of spin one-half 
constituents, and the methods of \cite{gh,fa} do not admit straightforward 
generalization for the case of the nuclei with other spins.  So, it was not 
clear from these works if the spin one-half formulae for the heavy particle 
gyromagnetic ratio may be applied without alteration for  
nuclei with other spins, for example in the case of deuterium.  This problem 
was resolved by quite general and abstract considerations in \cite{co} based 
on the representation theory of the Poincare group. It was shown there that 
respective expressions are equally valid for any spin of the nucleus. This 
problem is of more than academic interest since the high precision result 
for the ratio of the bound electron and bound deuteron gyromagnetic ratios 
was used in the 1986 adjustment of the fundamental constants \cite{ct} for 
extracting a high precision value of the free deuteron gyromagnetic ratio.

Recently there emerged a new interest in calculating binding corrections to 
the electron magnetic moment in high $Z$ atoms without expansion in 
$Z\alpha$.  The first calculations produced results (see, e.g., review in 
\cite{lpss}) which in the small $Z$ limit seemed to contradict the old 
results in \cite{gh,fa,co} although the results of later work \cite{bcs} 
are consistent with the \cite{gh,fa,co}.  It also turned out that the 
validity of the old results for arbitrary spin of the nuclei, which was 
proved in \cite{co} was not widely known\footnote{This question was raised 
in private communication by Profs.  B. Taylor and P.  Mohr. We have to admit 
that we ourselves have realized that Ref.\cite{co} contains the result for 
arbitrary spin only after completion of this note.}.  Inspired by these 
questions we decided to reconsider the problem once more.  We have studied a 
new approach to the problem which is different from all other methods used 
earlier \cite{gh,fa,co}. This approach is physically quite transparent 
and immediately provides results valid for particles with any spin with 
accuracy of order $(Z\alpha)^2$. It also clearly demonstrates that the 
problem of the proper description of the lowest order binding corrections to 
the gyromagnetic ratios for particles of any spin may be completely resolved 
in the nonrelativistic framework, and is not logically connected with the 
problem of Lorentz invariant description of bound states (see especially 
comments on the recoil corrections below).  We present below this new 
calculation of the gyromagnetic ratios for bound particles. 

\section{Spin Motion in an External Field}

In order to establish notation we start with the standard nonrelativistic 
Hamiltonian for the interaction of the magnetic moment with an external 
magnetic field

\beq
{\cal H}_{nonrel}=-{\mbox{\boldmath$\mu H$}},
\eeq

where the magnetic moment for arbitrary spin  in terms of its "native"
magneton $\mu_0=e/(2m)$\footnote{We use the system of units where 
$\hbar=c=1$, and use the definition above even in the case of negatively 
charged electron in order to be able to present a derivation of the binding 
corrections to the gyromagnetic ratio which is equally valid both for 
positively and negatively charged particles.}, and gyromagnetic ratio $g_s$, 
has the form

\beq
{\mbox{\boldmath$\mu$}}=g_s\mu_0{\bbox S}.
\eeq

Leading binding corrections of order $(Z\alpha)^2$ to the gyromagnetic 
ratio are of relativistic nature, and are induced by the relativistic 
corrections of order $v^2/c^2$ to the Hamiltonian. The observation that 
all terms of order $v^2/c^2$ in the spin Hamiltonian in an external field, 
may easily be restored with the help of the Bargmann-Michel-Telegdi (BMT) 
equation \cite{bmt}, is of crucial importance for the considerations below.

Consider four-vector $a_\mu$,  which is a relativistic generalization 
of the average spin vector in the rest frame \cite{blp}. The classical 
relativistic BMT equation describes the behavior of this four-vector  in an 
external electromagnetic field. This equation, which is a direct 
consequence of relativistic invariance, is equally valid for arbitrary 
magnitude of the average spin, and in terms of the particle gyromagnetic 
ratio has the form \cite{blp}

\beq
\frac{da^\mu}{d\tau}=\mu_0[g_sF^{\mu\nu}a_\nu
-(g_s-2)u^\mu F^{\nu\lambda}u_\nu a_\lambda],
\eeq

where $\tau$ is the relativistic proper time for the particle.

We would like to mention in passing that this equation nicely demonstrates 
the special role of the "normal" gyromagnetic moment $g_s=2$, which for any 
spin has a simple kinematic origin and is connected with the form of the 
ordinary Lorentz force (the famous Thomas one-half). Technically the special 
role of $g_s=2$ is connected with the correlation between the 
magnitude of the coefficient $e/m$ before the term with the magnetic field 
in the expression for the Lorentz force and the magnitude of the factor  
before the magnetic field in the nonrelativistic spin Hamiltonian above.

In terms of the ordinary time $t$ and the average spin three-vector 
\mbox{\boldmath$\zeta$} the BMT equation has a more complicated  form 
\cite{blp}

\beq   \label{bmt}
\frac{d\mbox{\boldmath$\zeta$}}{dt}=\mu_0\{\frac{g_sm+(g_s-2)(E_p-m)}{E_p}
[\mbox{\boldmath$\zeta\times H$}]+\frac{(g_s-2)E_p}{E_p+m}\mbox{(\boldmath$v 
H$}) [\mbox{\boldmath$v\times\zeta$}] 
\eeq 
\[ 
+\frac{g_sm+(g_s-2)E_p}{E_p+m}\mbox{\boldmath $[\zeta\times[E\times 
v]]$}\}.  
\]

In order to obtain the quantum mechanical description of the spin motion in 
the external field we follow the idea of \cite{blp}. First, we rewrite 
the BMT equation in terms of the canonical variables substituting ${\bbox 
v}=({\bbox p}-e{\bbox A})/E_p$. Then one may easily restore the quantum 
mechanical Hamiltonian which according to the canonical commutation 
relations leads to the BMT equation 

\beq         \label{qmh}
{\cal H}_{spin}={\cal H}'-\mu_0\frac{g_sm+(g_s-2)(E_p-m)}{E_p}{\bbox 
SH}+\mu_0\frac{(g_s-2)E_p}{(E_p+m)E_p^2}\left(({\bbox p}-e{\bbox A}) 
{\bbox H}\right)\left(({\bbox p}-e{\bbox A}) {\bbox S}\right)
\eeq 
\[ 
-\mu_0\frac{g_sm+(g_s-2)E_p}{(E_p+m)E_p}({\bbox S}
[{\bbox E}\times({\bbox p}-e{\bbox A})]), 
\]

where $H'$ contains all terms of the Hamiltonian which are spin independent. 

All relativistic corrections both to the normal and anomalous magnetic 
moments may be calculated with the help of the quantum mechanical 
Hamiltonian in \eq{qmh}. 

\subsection{Nonrecoil Limit}

Let us ignore first all recoil factors and calculate corrections to the 
gyromagnetic ratio of order $(Z\alpha)^2$. Physically this corresponds to a 
problem with a light particle in the field of an infinitely heavy Coulomb 
center. In this case canonical coordinates of the light particle coincide 
with the coordinates in the center of mass frame, and the problem 
simplifies.  We expand all coefficients in \eq{qmh} up to order ${\bbox 
p}^2/m^2$ and preserve only terms linear in the external magnetic field and 
spin

\beq 
{\cal H}_{exp}={\cal H}'-\mu_0\{g_s(1-\frac{{\bbox p}^2}{2m^2}){\bbox {SH}}
+(g_s-2)\frac{{\bbox p}^2}{2m^2}{\bbox {SH}}
-\frac{g_s-2}{2m^2}\mbox{(\boldmath$p 
H$)(\boldmath$p S$)} 
\eeq 
\[ 
-\frac{e}{m}[\frac{g_s}{2}+\frac{g_s-2}{2}]
(\mbox{\boldmath $S[E\times A]$})\}.  
\]

Next we simplify the perturbation, anticipating that the matrix elements 
will be calculated between spherically symmetric wave functions and that the 
external Coulomb electric field corresponds to the potential $V=-4\pi Ze/r$,

\beq
{\cal H}_{exp}=
{\cal H}'-\mu_0\{g_s[1-\frac{{\bbox p}^2}{2m^2}+\frac{Z\alpha}{6mr}]{\bbox 
{SH}} +(g_s-2)[\frac{{\bbox p}^2}{3m^2}+\frac{Z\alpha}{6mr}]{\bbox {SH}}\}.  
\eeq

Matrix elements between the Coulomb-Schrodinger wave functions for the 
$nS$ states are given by the relations

\beq
<n|\frac{Z\alpha}{mr}|n>=<n|\frac{{\bbox p}^2}{m^2}|n>
=\frac{(Z\alpha)^2}{n^2},
\eeq

and we easily obtain

\beq
<n|{\cal H}_{exp}|n>=<n|{\cal H}'|n>-\mu_0\{g_s(1-\frac{(Z\alpha)^2}{3n^2})
+(g_s-2)\frac{(Z\alpha)^2}{2n^2}\}{\bbox {SH}},
\eeq

or

\beq
g_{bound}
=g_s[(1-\frac{(Z\alpha)^2}{3n^2})
+\frac{g_s-2}{g_s}\frac{(Z\alpha)^2}{2n^2}]
\approx g_s[(1-\frac{(Z\alpha)^2}{3n^2})
+\frac{\alpha(Z\alpha)^2}{4\pi n^2}].
\eeq

This result reproduces the old results \cite{gh,fa,co} in the 
nonrecoil approximation. 

\section{Center of Mass Motion in External Field}

Consider next the Coulomb bound system of two particles. We are seeking 
corrections to the magnetic moments of the constituents, and the first 
problem is to separate center of mass motion from the internal degrees of 
freedom. This is of course trivial in the absence of an external field 
due to translation invariance. But an external field breaks translation 
invariance and the usual variables in the respective Schrodinger equation  
do not separate any more, even in the nonrelativistic 
case.  One might think that this is completely unimportant, since in any 
case we consider the external field as a small perturbation and are 
interested only in the terms which are linear in the external field. 
However, as we will see below, even in the case of vanishingly small 
external field one cannot ignore its influence on the proper description of 
the center of mass motion if recoil corrections are to be accounted for.

Let us consider two charged nonrelativistic particles which interact via the
Coulomb potential in an external homogeneous magnetic field. The Lagrangian 
of this system has the form

\beq
L=\frac{m_i \dot{{\bbox r}}_i^2}{2}+ e_i{{\bbox A}_i\dot{\bbox 
r}_i}-V({\bbox r}_1-{\bbox r}_2), 
\eeq

where ${\bbox A}_i={\bbox H}\times {\bbox r}_i/2$ and $V({\bbox 
r}_1-{\bbox r}_2)=4\pi e_1e_2/|{\bbox r}_1-{\bbox r}_2|$.  The center of 
mass and relative coordinates are defined by the standard relations

\beq
{\bbox r}={\bbox r}_1-{\bbox r}_2,
\eeq
\[
{\bbox R}=\mu_1{\bbox r}_1+\mu_2{\bbox r}_2,
\]

where $\mu_1=m_1/(m_1+m_2)$, $\mu_2=m_2/(m_1+m_2)$. In these variables 
the Lagrangian has the form

\beq
L=\frac{(m_1+m_2)\dot{{\bbox R}}^2}{2}+\frac{m_r
\dot{{\bbox r}}^2}{2}+(e_1+e_2){\bbox A}({\bbox R})\dot{{\bbox R}} 
\eeq
\[
+\frac{e_1\mu_2-e_2\mu_1}{2}(
[{\bbox H}\times {\bbox r}]\dot{{\bbox R}}+[{\bbox H}\times{\bbox  
R}]\dot{{\bbox r} })+ (e_1\mu_2^2+e_2\mu_1^2){\bbox A}({\bbox 
r})\dot{{\bbox r}}-V({\bbox r}).  
\]

Respective canonical momenta are as follows

\beq
{\bbox P}=\frac{\partial L}{\partial\dot{\bbox R}}=(m_1+m_2)\dot{\bbox 
R}+(e_1+e_2){\bbox A}({\bbox R})+(e_1\mu_2-e_2\mu_1){\bbox A}({\bbox r}) 
\eeq 
\[ 
={\bbox p}_1+{\bbox p}_2, 
\] 
\[ 
{\bbox p}=\frac{\partial L}{\partial\dot{\bbox r}}
=m_r{\dot{\bbox r}} +(e_1\mu_2^2+e_2\mu_1^2){\bbox A}({\bbox r})
+(e_1\mu_2-e_2\mu_1){\bbox A}({\bbox R}) 
\] 
\[ 
=\frac{\mu_2-\mu_1}{2}{\bbox P} 
+\frac{{\bbox p}_1-{\bbox p}_2}{2}, 
\] 
\[ 
{\bbox p}_1=\mu_1{\bbox P}+{\bbox p}, 
\] 
\[ 
{\bbox p}_2=\mu_2{\bbox P}-{\bbox p}, 
\]

and the Hamiltonian is equal to

\beq            \label{freeham}
{\cal H}=\frac{({\bbox p}_1-e_1{\bbox A}_1)^2}{2m_1}+\frac{({\bbox 
p}_2-e_2{\bbox A}_2)^2}{2m_2} 
\eeq 
\[ 
=\frac{({\bbox P}-(e_1+e_2){\bbox A}({\bbox R}))^2}{2(m_1+m_2)} 
+\frac{({\bbox p}-(e_1\mu_2^2+e_2\mu_1^2){\bbox A}({\bbox r}))^2}{2m_r}
\]
\[
+(e_1\mu_2-e_2\mu_1)^2\frac{{\bbox A}^2({\bbox R})}{2m_r}
+(e_1\mu_2-e_2\mu_1)^2\frac{{\bbox A}^2({\bbox r})}{2(m_1+m_2)}
\]
\[
+(e_1\mu_2-e_2\mu_1)[(e_1\mu_2+e_2\mu_1)\frac{\bbox {A(r)A(R)}}{m_r} 
-\frac{\bbox {A(R)p}}{m_r}-\frac{\bbox {A(r)P}}{m_1+m_2}]+V({\bbox r}).
\]

Naively, one would expect that this Hamiltonian the describes motion of the 
bound system with canonical coordinates $\bbox P,\bbox R$ in the external 
field as motion of a free particle with charge $(e_1+e_2)$ and mass 
$m_1+m_2$.  However, this is far from evident, since the variables 
are not separated in this Hamiltonian.  We need an additional principle to 
confirm that this Hamiltonian properly describes the bound system as a 
whole.  It is not too difficult to discover such a principle.  It is well 
known that the momentum a free particle in the external magnetic field does 
not commute with the Hamiltonian, and instead one should have the 
commutation relation (see, e.g., \cite{gh})

\beq      \label{freecom}
[{\cal H},{\bbox P}+(e_1+e_2){\bbox A}({\bbox R})]=0.
\eeq

For the Hamiltonian in \eq{freeham} one obtains

\beq
[{\cal H},{\bbox P}+(e_1+e_2){\bbox A}({\bbox R})]
=-i(e_1\mu_2-e_2\mu_1)\frac{[\left({\bbox p}- 
(e_1\mu_2+e_2\mu_1){\bbox A}({\bbox r})\right)\times {\bbox 
H}]}{m_r}+O({\bbox H}^2).  
\eeq

This commutator, of course, vanishes in the absence of an external field, 
but it is nonvanishing in any external field and its value is not even
suppressed by the mass ratio when one of the masses becomes much larger than 
another. Hence, the Hamiltonian in \eq{freeham} does not appears to describe 
the center of mass motion as propagation of a free particle in an external 
magnetic field, and its dependence on the internal coordinate cannot be used 
for calculation of any properties of the bound system, in particular its 
gyromagnetic ratio.

The Hamiltonian in \eq{freeham} still admits a conserved vector 
${\bbox P}+(e_1+e_2){\bbox A}({\bbox R})+(e_1\mu_2-e_2\mu_1){\bbox 
A}({\bbox r})$,

\beq
[{\cal H},{\bbox P}+(e_1+e_2){\bbox A}({\bbox R})+(e_1\mu_2-e_2\mu_1){\bbox 
A}({\bbox r})]=0.  
\eeq

This vector is equal to the sum of the respective conserved vectors for the 
free constituents 
${\bbox P}+(e_1+e_2){\bbox 
A}({\bbox R})+(e_1\mu_2-e_2\mu_1){\bbox A}({\bbox r})={\bbox p}_1+e_1{\bbox 
A}({\bbox r}_1)+{\bbox p}_2+e_2{\bbox A}({\bbox r}_2)$, and it is conserved 
because addition to the free Hamiltonian of the Coulomb potential which 
depends only on the relative distance does not break conservation of this 
sum.

Now it is clear that the proper description of the center of mass 
motion may be achieved with the help of a unitary transformation which 
transforms the conserved vector into the one corresponding to a free charged 
particle ${\bbox P}+(e_1+e_2){\bbox A}({\bbox R})+(e_1\mu_2-e_2\mu_1){\bbox 
A}({\bbox r})\rightarrow {\bbox P}+(e_1+e_2){\bbox A}({\bbox R})$. This 
recipe was suggested in \cite{gh}, and the respective unitary transformation 
has the form

\beq
U=e^{-i(e_1\mu_2-e_2\mu_1){\bbox A}({\bbox r}){\bbox R}}.
\eeq

The unitary transformed Hamiltonian

\beq
{\cal H}'=U^{-1}{\cal H}U
=\frac{({\bbox P}-(e_1+e_2){\bbox A}({\bbox R}))^2}{2(m_1+m_2)}
+\frac{({\bbox p}-(e_1\mu_2^2+e_2\mu_1^2){\bbox A}({\bbox r}))^2}{2m_r}
\eeq
\[
+2(e_1\mu_2-e_2\mu_1)^2{\bbox A}({\bbox r})\frac{{\bbox P}-(e_1+e_2){\bbox 
A}({\bbox R})}{m_1+m_2} +2(e_1\mu_2-e_2\mu_1)^2\frac{{\bbox A}^2({\bbox 
r})}{(m_1+m_2)}, 
\]

now satisfies the commutation relation

\beq      
[{\cal H}',{\bbox P}+(e_1+e_2){\bbox A}({\bbox R})]=0,
\eeq

characteristic for a free particle of charge $e_1+e_2$ in an 
external field (compare \eq{freecom}). Hence, in all calculations of 
the properties of the bound system one should use the unitary transformed 
Hamiltonian. Any perturbations which we will add below to the Hamiltonian 
also should be unitary transformed.

\section{Recoil Corrections}

Unitary transformation of the spin interaction Hamiltonian corresponding to 
the BMT equation reduces to substitution of the transformed 
momenta instead of respective particle velocities in \eq{bmt}

\beq
{\bbox v}_{i}\rightarrow {\bbox p}'_i-e_i{\bbox 
A}({\bbox r}_i)=\mu_i[{\bbox P}-(e_1+e_2){\bbox A}({\bbox R})]\pm 
[{\bbox p}-(e_i-(e_1+e_2)\mu_i^2]{\bbox A}({\bbox r})), 
\eeq

and the spin Hamiltonian acquires (say, for the first particle) the form

\beq         
{\cal H}_{spin}=-\mu_0\frac{g_sm_1+(g_s-2)(E_{p'_1}-m_1)}
{E_{p'_1}}{\bbox SH}
\eeq 
\[
+\mu_0\frac{(g_s-2)E_{p'_1}}{(E_{p'_1}+m_1)E_{ p'_1}^2}
(({\bbox p}'_1-e_1{\bbox A}({\bbox r}_1)) {\bbox H})(( 
{\bbox p}'_1-e_1{\bbox A}({\bbox r}_1)){\bbox S}) 
\]
\[ 
-\mu_0\frac{g_sm_1+(g_s-2)E_{p'_1}}{(E_{p'_1}+m_1)
E_{p'_1}}({\bbox S}[{\bbox E}\times({\bbox p}'_1-e_1{\bbox 
A}({\bbox r}_1))]).  
\]

Thus far  we completely ignored all interaction  terms between the two 
particles except the Coulomb term.  This is a valid approximation in the 
nonrecoil limit, but it is easy to realize that the Breit interaction (the 
transverse photon exchange) immediately generates terms linear and quadratic 
in the mass ratio. Hence, we have to amend the Hamiltonian above by 
the Breit interaction term.  This term may be easily calculated in the first 
order perturbation theory (see, e.g., \cite{gh}) and has the form (again 
after the unitary transformation)

\beq
{\cal H}_{Br}=g_s\frac{4\pi e_1e_2}{2m_1m_2}{\bbox S}
[\frac{\bbox r}{r^3}\times({\bbox p}'_2-e_2{\bbox A}({\bbox r}_2))]
=g_s\mu_0\frac{4\pi e_2}{m_2}{\bbox S}[\frac{\bbox r}{r^3}\times
({\bbox p}'_2-e_2{\bbox A}({\bbox r}_2))],
\eeq

which contains a term linear in $\bbox H$

\beq
{\cal H}_{Br}\approx g_s\mu_0\frac{[e_2-(e_1+e_2)\mu_2^2]}{m_2}{\bbox 
S}[{\bbox E}\times {\bbox A}({\bbox r})] =g_s\mu_0\frac{4\pi 
e_2[e_2-(e_1+e_2)\mu_2^2]}{3m_2r}{\bbox {SH}}.  
\eeq

Now we add the Breit interaction term to the spin Hamiltonian 
above, substitute the center of mass and relative coordinates, go to the 
center of mass system, preserve only the terms of order $(Z\alpha)^2$ and 
obtain

\beq         
{\cal H}_{tot}
=-\mu_0({\bbox{ SH}})\{g_s(1-\frac{\bbox p^2}{2m_1^2}) +(g_s-2)
\frac{\bbox p^2}{2m_1^2}
-\frac{g_s-2}{2}\frac{\bbox p^2}{3m_1^2} 
\eeq 
\[ 
-\frac{4\pi e_2[e_1-(e_1+e_2)\mu_1^2]}{3m_1r}[\frac{g_s}{2}+\frac{g_s-2}{2}]
-g_s\frac{4\pi e_2[e_2-(e_1+e_2)\mu_2^2]}{3m_2r}\}.
\]

Then 

\beq
g_{1bound}=g_s\{(1-\frac{<\bbox p^2>}{2m_1^2})
-[\frac{4\pi e_2[e_1-(e_1+e_2)\mu_1^2]}{6m_1}
+\frac{4\pi e_2[e_2-(e_1+e_2)\mu_2^2]}{3m_2}]<\frac{1}{r}>\}
\eeq
\[
+(g_s-2)\{\frac{<\bbox p^2>}{2m_1^2}-\frac{<\bbox p^2>}{6m_1^2}-\frac{4\pi 
e_2[e_1-(e_1+e_2)\mu_1^2]}{6m_1}<\frac{1}{r}>\}.
\]

Now we use matrix elements

\beq
<n|\frac{1}{r}|n>
=-\frac{4\pi e_1e_2}{n^2}{m_r}\approx
-\frac{4\pi e_1e_2}{n^2}(1-\frac{m_1}{m_2}+(\frac{m_1}{m_2})^2),
\eeq
\[
<n|\frac{\bbox p^2}{m_1^2}|n>
=\frac{(4\pi e_1e_2)^2}{n^2}\mu_2^2\approx
\frac{(4\pi e_1e_2)^2}{n^2}(1-2\frac{m_1}{m_2}+3(\frac{m_1}{m_2})^2),
\]

and obtain

\beq
g_{1bound}=
g_s\{(1-\frac{(4\pi e_1e_2)^2}{2n^2}\mu_2^2)
+[\frac{4\pi e_2\mu_2[e_1-(e_1+e_2)\mu_1^2]}{6}
\eeq
\[
+\frac{4\pi e_2\mu_1[e_2-(e_1+e_2)\mu_2^2]}{3}]\frac{4\pi e_1e_2}{n^2}\}
\]
\[
+(g_s-2)\{\frac{(4\pi e_1e_2)^2}{3n^2}\mu_2^2
+\frac{4\pi e_2\mu_2[e_1-(e_1+e_2)\mu_1^2]}{6}\frac{4\pi 
e_1e_2}{n^2}\}.  
\]

Our treatment was completely symmetric with respect to both constituents, so 
the gyromagnetic ratio for the second particle may be obtained from the 
expression above by a simple substitution of indices

\beq
g_{2bound}=g_s\{(1-\frac{(4\pi e_1e_2)^2}{2n^2}\mu_1^2)
+[\frac{4\pi e_1\mu_1[e_2-(e_1+e_2)\mu_2^2]}{6}
\eeq
\[
+\frac{4\pi e_1\mu_2[e_1-(e_1+e_2)\mu_1^2]}{3}]\frac{4\pi e_1e_2}{n^2}\}
+(g_s-2)\{\frac{(4\pi e_1e_2)^2}{3n^2}\mu_1^2
\]
\[
+\frac{4\pi e_1\mu_1[e_2-(e_1+e_2)\mu_2^2]}{6}\frac{4\pi 
e_1e_2}{n^2}\}.  
\]

It is interesting to consider the formulae for the gyromagnetic ratios of 
the constituents in the case when one particle (say, the second one) is much 
heavier than the other. It is physically evident that in the linear 
approximation in the small mass ratio one may easily obtain the binding 
corrections to the gyromagnetic ratios from the nonrecoil treatment in the 
second part above, simply by substituting reduced mass instead of the 
particle mass in the expressions for the matrix elements and adding the 
Breit contribution which is linear in the small mass ratio from the very 
beginning. We have checked that this simple-minded approach reproduces the 
formulae above with linear accuracy in the mass ratio. On the other hand 
naive calculation of the terms linear in the mass ratio from the two 
particle Hamiltonian above, ignoring the unitary transformation, leads to a 
wrong result. This remark emphasizes once more the importance of the unitary 
transformation above for the proper separation of the center of mass motion. 
Let us emphasize that the need for the unitary transformation emerges 
already in a completely nonrelativistic framework, and ignoring it 
leads to incorrect results even in the linear approximation in the small 
mass ratio.

\section{Conclusion}

In this short note we proposed a new method for calculation of the 
leading binding corrections to the gyromagnetic ratios of bound particles. 
We have confirmed the validity of the old results by using the BMT equation 
to obtain the magnetic interactions in the two-particle Hamiltonian.
Our new approach clearly demonstrates the purely kinematic nature of the 
distinction between the normal and anomalous parts of the particle
magnetic moment, and the independence of the leading binding corrections on 
the magnitude of the particle spin.

\acknowledgements
We would like to thank Barry Taylor and Peter Mohr who attracted our 
attention to the problem discussed in this note.

M. E. is deeply grateful for the kind hospitality of the Physics 
Department at Penn State University, where this work was performed. The 
authors appreciate the support of this work by the National Science 
Foundation under grant number PHY-9421408.


\begin{thebibliography}{99}

\bibitem{gh} H. Grotch and R. A. Hegstrom, Phys. Rev. A {\bf 4} (1971), 59.  

\bibitem{fa} R. N. Faustov, Phys. Lett. B {\bf 33} (1970), 422.  

\bibitem{co} F. E. Close and H. Osborn, Phys. Lett. B {\bf 34} (1971), 400.  

\bibitem{ct} E. R. Cohen and B. N. Taylor, Rev. Mod. Phys. {\bf 59} (1987), 
1121.  

\bibitem{lpss} I. Lindgren et al, {\it in} "Proc. of the 15th International 
Conference on Atomic Physics", Amsterdam, 1996.

\bibitem{bcs} S. A. Blundell, K. T. Cheng, and J. Sapirstein, Phys. Rev. A 
{\bf 55} (1997), 1857.

\bibitem{bmt} V. Bargmann, L. Michel, and V. Telegdi, Phys. Rev. Lett., {\bf 
2} (1959), 435.

\bibitem{blp} V. B. Berestetskii, E. M. Lifshitz, and L. P. Pitaevskii, 
"Quantum electrodynamics", 2nd Edition, Pergamon Press, Oxford, 1982.

\end{thebibliography}
\end{document}